\documentclass[12pt]{article}
\usepackage{amsmath,amssymb,amsthm,amsxtra,bbm,overpic,bm,epsfig,subfigure}
\usepackage[colorlinks]{hyperref}
\usepackage{mathrsfs}
\usepackage{enumitem}
\usepackage{graphicx}
\usepackage{color}
\usepackage{comment}
\usepackage{epstopdf}
\usepackage{float}
\usepackage{cite}
\usepackage{upgreek,overpic,setspace,multirow,textcomp,orcidlink}
\numberwithin{equation}{section}
\allowdisplaybreaks[2]

\usepackage{slashed,stmaryrd}
\usepackage{subcaption}

\def\thefootnote{\fnsymbol{footnote}}

\usepackage{lscape}%
\usepackage{array}
\usepackage{booktabs}%

\begin{document}
	
	\vspace{0.2cm}
	
	\begin{center}
        {\Large\bf Effects of the Matter Potential at One-Loop Level on Neutrino Oscillations in Long-Baseline Experiments}
	\end{center}
	
	\vspace{0.2cm}
	
	\begin{center}
		{\bf Jihong Huang}{\orcidlink{0000-0002-5092-7002}},$^{1,2}$\footnote{E-mail: huangjh@ihep.ac.cn}
		\quad
		{\bf Tommy Ohlsson}{\orcidlink{0000-0002-3525-8349}},$^{3,4}$\footnote{E-mail: tohlsson@kth.se}
		\quad
		{\bf Sampsa Vihonen}{\orcidlink{0000-0001-7761-2847}},$^{3,4}$\footnote{E-mail: vihonen@kth.se}
		\quad
		{\bf Shun Zhou}{\orcidlink{0000-0003-4572-9666}},$^{1,2}$\footnote{E-mail: zhoush@ihep.ac.cn}
		\\
		\vspace{0.2cm}
		{\small
			$^{1}$Institute of High Energy Physics, Chinese Academy of Sciences, Beijing 100049, China\\
			$^{2}$School of Physical Sciences, University of Chinese Academy of Sciences, Beijing 100049, China\\
                $^{3}$Department of Physics, School of Engineering Sciences, KTH Royal Institute of Technology,\\
                AlbaNova University Center, Roslagstullsbacken 21, SE–106 91 Stockholm, Sweden\\
                $^{4}$The Oskar Klein Centre, AlbaNova University Center, Roslagstullsbacken 21,\\
                SE–106 91 Stockholm, Sweden}
	\end{center}

	\vspace{0.5cm}
	
	\begin{abstract}
		In this work, we investigate in a quantitative way how much radiative corrections to the matter potential for neutrino oscillations can impact the sensitivity to neutrino mass ordering in long-baseline accelerator experiments. Using numerical simulations for the future experiment DUNE, we find that the statistical significance for excluding the incorrect mass ordering can be enhanced by about $0.4\sigma$ if a one-loop correction of $2.0\%$---based on the Fermi coupling constant $G^{}_\mu$ derived from measurements of muon lifetime---is included. The radiative corrections at one-loop level lead to resolving the neutrino mass ordering at $5\sigma$ confidence level 4–9 days earlier than at tree level. In contrast, the sensitivity to leptonic CP violation in DUNE is essentially unchanged. Finally, we emphasize that one-loop corrections should be incorporated into analyses of future neutrino oscillation data in a consistent and systematic manner.
	\end{abstract}
    
	\def\thefootnote{\arabic{footnote}}
	\setcounter{footnote}{0}
	
	\newpage
	
	\section{Introduction}
	
	The experimental discoveries of atmospheric, solar, reactor, and accelerator neutrino oscillations have established that neutrinos are massive and leptonic flavor mixing is significant~\cite{ParticleDataGroup:2024cfk,Xing:2020ijf}. The next-generation of neutrino oscillation experiments aims to determine whether three neutrino masses $m_1, m_2$, and $m_3$ follow the normal mass ordering (NO, {\em i.e.}, $m_1^{} < m_2^{} < m_3^{}$) or the inverted mass ordering (IO, {\em i.e.}, $m_3^{} < m_1^{} < m_2^{}$) and to search for CP violation in the lepton sector. Various experiments have been designed to resolve the neutrino mass ordering. One is the medium-baseline reactor neutrino experiment JUNO, which is planned to start taking data later this year~\cite{JUNO:2015zny}. Another is the future long-baseline accelerator neutrino experiment DUNE~\cite{DUNE:2020ypp}, which is currently under construction. While JUNO will probe the neutrino mass ordering by utilizing the interference between oscillations driven by the two mass-squared differences $\Delta m_{21}^2 \equiv m_2^2 - m_1^2$ and $\Delta m_{31}^2 \equiv m_{3}^2 - m_{1}^2$, DUNE will make use of matter effects to determine the sign of $\Delta m_{31}^2$. The neutrino mass ordering can also be tested in other types of experiments, see {\em e.g.}~Ref.~\cite{DeSalas:2018rby} for a comprehensive review.
    
    Matter effects~\cite{Wolfenstein:1977ue, Wolfenstein:1979ni, Mikheyev:1985zog, Mikheev:1986wj}, caused by coherent forward scattering off ordinary matter, can be described by the matter potential (sometimes named the Wolfenstein potential)~\cite{Wolfenstein:1977ue}. In the Standard Model, the matter potential at tree level has been derived nearly fifty years ago, while the complete calculation at one-loop level in the on-shell renormalization scheme has recently been performed in Ref.~\cite{Huang:2023nqf}. With the electromagnetic fine-structure constant $\alpha$, the gauge-boson masses $m_W^{}$ and $m_Z^{}$, the Higgs-boson mass $m_h^{}$, and the on-shell masses of the Standard Model fermions $m_f^{}$ chosen as input parameters, it has been found that there is a $5.8\%$ correction to the charged-current (CC) potential and that to the neutral-current (NC) potential is about $8.2\%$. Since the NC potential is the same for all neutrino flavors, only the CC potential affects the oscillation behavior. Now, it is quite interesting to discuss how significant such corrections are in determining the mass ordering and measuring the CP-violating phase in future experiments. The motivation for this work is two-fold. First, in the precision era of neutrino physics, oscillation parameters can be measured even to the sub-percent level~\cite{JUNO:2022mxj,Capozzi:2025wyn}. Therefore, the effects of radiative corrections with the same order of magnitude cannot be ignored. For the detection of solar neutrinos, the one-loop correction to the cross-section for neutrino-electron scattering has been considered in Ref.~\cite{Bahcall:1995mm}. As for the matter potential, the inclusion of one-loop corrections renders its input value closer to the true one, and thus helps to improve the experimental accuracy of the oscillation parameters. Second, the uncertainty in the matter density is generally assumed to be $5\%$ and can be significantly reduced for specific long-baseline accelerator neutrino experiments. For example, there is an uncertainty of $2\%$ in the matter density along the baseline of DUNE~\cite{DUNE:2021cuw}. Since the uncertainty in the matter density is comparable to or even smaller than the one-loop corrections, it is necessary to take into account the latter in analyses of neutrino oscillation data from long-baseline experiments.
    
    In this work, we investigate the impact of such radiative corrections on DUNE in detail. We simulate the neutrino events in the $\nu_\mu^{} \to \nu_e^{}$ appearance channel with the tree- and loop-level matter potentials, and make a comparison between the event spectra for these two cases. The constant approximation and the Shen-Ritzwoller profile~\cite{Shen:2016kxw} of the matter density, together with their corresponding uncertainties, are both considered in the simulations. Then, we study the impact on determining the neutrino mass ordering and measuring the leptonic CP-violating phase. We find that the radiative corrections could increase the sensitivity of DUNE to the mass ordering by about $0.4\sigma$ confidence level (CL), while there is no notable influence on the measurement of the CP-violating phase.  
	
	The remaining parts of this work are organized as follows. In Sec.~\ref{sec:nu_osc_matter}, we present a general discussion on neutrino oscillations in matter, including the tree-level matter potential, its one-loop corrections, and the impact of the matter density profile and its uncertainties on neutrino oscillations. Then, in Sec.~\ref{sec:num}, the numerical analysis of neutrino oscillations in DUNE is performed, and the effects of the one-loop corrections on experimental sensitivities to the neutrino mass ordering and the leptonic CP-violating phase are also discussed. Finally, in Sec.~\ref{sec:sum}, our main results and conclusions are summarized.
	
    \section{Neutrino Oscillations in Matter}
    
    \label{sec:nu_osc_matter}

    In the framework of three-flavor neutrino oscillations, leptonic flavor mixing is described by the Pontecorvo-Maki-Nakagawa-Sakata matrix $U$~\cite{Pontecorvo:1957cp,Maki:1962mu}, which can be expressed in the standard parametrization as~\cite{ParticleDataGroup:2024cfk}
    \begin{eqnarray}
        U = \begin{pmatrix}
			c_{12}^{} c_{13}^{} & s_{12}^{} c_{13}^{} & s_{13}^{} {\rm e}^{-{\rm i} \delta_{\rm CP}^{}}  \\
			-s_{12}^{} c_{23}^{} - c_{12}^{} s_{13}^{} s_{23}^{} {\rm e}^{{\rm i}\delta_{\rm CP}^{}} & c_{12}^{} c_{23}^{} - s_{12}^{} s_{13}^{} s_{23}^{} {\rm e}^{{\rm i} \delta_{\rm CP}^{}} & c_{13}^{} s_{23}^{} \\
			s_{12}^{} s_{23}^{} - c_{12}^{} s_{13}^{} c_{23}^{} {\rm e}^{{\rm i}\delta_{\rm CP}^{}} & -c_{12}^{} s_{23}^{} - s_{12}^{} s_{13}^{} c_{23}^{} {\rm e}^{{\rm i} \delta_{\rm CP}^{}} & c_{13}^{} c_{23}^{} 
		\end{pmatrix} \;,
    \end{eqnarray}
    where $s_{ij}^{} \equiv \sin \theta_{ij}^{}$ and $c_{ij}^{} \equiv \cos \theta_{ij}^{}$ with $\theta^{}_{ij}$ (for $i,j = 1,2; 1,3; 2,3$) being the three mixing angles and $\delta_{\rm CP}^{}$ is the Dirac CP-violating phase. The two possible Majorana CP-violating phases are not considered in this work, since they are irrelevant to neutrino oscillations in both vacuum and matter. When expanded in terms of the small ratio of mass-squared differences $\Delta m_{21}^2 / \Delta m_{31}^2 \simeq 0.03$ and the smallest mixing angle $\sin^2 \theta_{13}^{} \simeq 0.02$, the oscillation probability in the appearance channel $\nu_\mu^{} \to \nu_e^{}$ for (accelerator) neutrino oscillation experiments with the baseline length $L$ and the neutrino energy $E_\nu$ can be approximated and written as~\cite{Cervera:2000kp,Freund:2001pn,Akhmedov:2004ny,Nunokawa:2007qh}
	\begin{eqnarray}
    \label{eq:numu_to_nune}
		P_{\mu e} \equiv P\left(\nu_{\mu}^{} \to \nu_{e}^{}\right) &\simeq& \sin^{2} \theta_{23}^{} \sin^{2} 2 \theta_{13}^{} \frac{\sin^{2}\left(\Delta_{31}^{} -a L\right)}{\left(\Delta_{31}^{} -a L\right)^{2}} \Delta_{31}^{2} \nonumber \\ 
		&& +\sin 2 \theta_{23}^{} \sin 2 \theta_{13}^{} \sin 2 \theta_{12}^{} \frac{\sin \left(\Delta_{31}^{} - a L\right)}{\left(\Delta_{31}^{}-a L\right)} \Delta_{31}^{} \frac{\sin (a L)}{(a L)} \Delta_{21}^{} \cos \left(\Delta_{31}^{}+\delta_{\rm CP}^{}\right) \nonumber \\ 
		&& +\cos^{2} \theta_{23}^{} \sin^{2} 2 \theta_{12}^{} \frac{\sin^{2}(a L)}{(a L)^{2}} \Delta_{21}^{2} \;,
	\end{eqnarray}
	where $\Delta_{ij}^{} \equiv \Delta m_{ij}^2 L/(4E_\nu)$ and the matter parameter $a \equiv {\cal V}_{\rm CC}^{}/2$ with ${\cal V}_{\rm CC}^{}$ being the CC matter potential. For the oscillation probability of antineutrinos $\overline{P}_{\mu e}^{} \equiv P\left(\overline{\nu}_{\mu}^{} \to \overline{\nu}_{e}^{}\right)$, one must replace $\delta_{\rm CP}^{} \to - \delta_{\rm CP}^{}$ and $a \to -a$. 
    
    As indicated by the first two terms on the right-hand side of Eq.~(\ref{eq:numu_to_nune}), the sign of $\Delta_{31}^{}$ relative to the matter parameter $a$ plays an important role. Furthermore, both $\Delta^{}_{31}$ in the NO case and $a L$ are positive, and thus, their difference $\Delta^{}_{31} - a L$ appears in the probability. However, $\Delta_{31}^{}$ changes its sign in the IO case, which means that $\sin(\Delta^{}_{31} - aL)/(\Delta^{}_{31} - aL) = \sin(-\Delta^{}_{31} + aL)/(-\Delta^{}_{31} + aL)$, where actually the sum $|\Delta^{}_{31}| + aL$ appears instead in the probability. Therefore, the oscillation probabilities in the NO and IO cases will be remarkably different and distinguishable, by which it can determine the neutrino mass ordering with the help of matter effects. For antineutrinos, the matter parameter $a$ changes sign. However, the feasibility of determining the neutrino mass ordering in long-baseline accelerator neutrino experiments can be understood in a similar manner.
    
	\subsection{Matter Potential and Its One-Loop Corrections}
    
    The amplitudes of the coherent forward scattering $\nu_\alpha^{} + f \to \nu_\alpha^{} + f$ for $\alpha=e,\mu,\tau$ and $f=u,d,e$ can be divided into NC and CC parts. Assuming a homogeneous and isotropic medium, the effective potentials can be obtained by averaging the effective Hamiltonian over all possible states of background fermions. At tree level, we obtain~\cite{Giunti:2007ry,Xing:2011zza}
	\begin{eqnarray}
		\label{eq:Veff}
		{\cal V}_{\rm NC}^{} = \frac{\pi \alpha m_Z^2}{m_W^2 \left(m_Z^2 - m_W^2\right)} N_f^{} c_{\rm V, NC}^f \;, \qquad {\cal V}_{\rm CC}^{} = \frac{\pi \alpha m_Z^2}{m_W^2 \left(m_Z^2 - m_W^2\right)} N_e^{} c^e_{\rm V, CC} \;,
	\end{eqnarray}
	where the fine-structure constant $\alpha$, the on-shell masses of $W$ and $Z$ bosons $m_W^{},m_Z^{}$ are chosen as input parameters, and $N^{}_f$ is the net number density of the background fermion $f$. Only the vector-type couplings $c^f_{\rm V,NC} \equiv I^3_f - 2 s^2 Q_f^{}$ and $c^e_{\rm V,CC} = 1$ are involved, where the quantity $s \equiv \sin\theta_{\rm w}^{}$, with the weak mixing angle $\theta_{\rm w}^{}$, and $I^3_f$ and $Q_f^{}$ are the weak isospin component and the electric charge of $f$, respectively. Accordingly, the matter potentials change sign for antineutrinos.

    \begin{table}[t]
    \centering
    \begin{spacing}{1.25}
        \begin{tabular}{c|c||c|c||c|c||c|c}
        \hline\hline
        $m_W^{}/{\rm GeV}$ & 80.369 & $m_e^{}/{\rm MeV}$ & 0.511 & $m_u^{}/{\rm MeV}$ & 62 & $m_d^{}/{\rm MeV}$ & 83 \\ \hline 
        $m_Z^{}/{\rm GeV}$ & 91.188 &  $m_\mu^{}/{\rm MeV}$  & 105.658 & $m_c^{}/{\rm GeV}$ & 1.67 & $m_s^{}/{\rm MeV}$ & 215 \\ \hline
        $m_h^{}/{\rm GeV}$ & 125.20 & $m_\tau^{}/{\rm GeV}$ & 1.777 & $m_t^{}/{\rm GeV}$ & 172.57 & $m_b^{}/{\rm GeV}$ & 4.78 \\ \hline\hline
        \end{tabular} 
    \caption{Input values of on-shell masses of gauge bosons, the Higgs boson and charged fermions~\cite{ParticleDataGroup:2024cfk}. See the main text for further discussion.}\label{tab:inputs}
        \end{spacing}
    \end{table}
    
    At one-loop level, the matter potential can be calculated by extracting the corrections to the vector-type couplings from the renormalized scattering amplitudes. Then, we arrive at the relative magnitude of the one-loop corrections to the NC couplings as $\Delta c^f_{\rm V,NC}/c^{f}_{\rm V,NC}$ with $\Delta c_{\rm V,NC}^f \equiv \widehat{c}_{\rm V,NC}^f - c_{\rm V,NC}^f$ being the differences between the one-loop couplings $\widehat{c}_{\rm V,NC}^f$ and the tree-level ones $c^f_{\rm V,NC}$. Similarly, the relative correction to the CC coupling of electrons is $\Delta c^e_{\rm V, CC}/c^e_{\rm V,CC}$ with $\Delta c_{\rm V,CC}^e \equiv \widehat{c}_{\rm V,CC}^e - c_{\rm V,CC}^e$. 
    
    In Ref.~\cite{Huang:2023nqf}, the explicit expressions for $\Delta c^f_{\rm V,NC}$ and $\Delta c^e_{\rm V, CC}$ are given, which are formulated in terms of the fine-structure constant $\alpha \simeq 1/137.035999084$ and the on-shell masses of gauge bosons, the Higgs boson, and charged fermions. Their numerical values are listed in Table~\ref{tab:inputs}. The masses of gauge bosons and heavy quarks ($c$, $b$, and $t$) are actually pole masses measured in experiments, while we refer to them as the on-shell masses, since the difference between those two sets of masses appears at two-loop level~\cite{Sirlin:1991fd,Sirlin:1991rt,Pilaftsis:1997dr} and can therefore be safely neglected. Due to the non-perturbative nature of strong interactions at low energies, light quarks ($u$, $d$, and $s$) are not physical degrees of freedom and their on-shell masses are not well defined. We refer to the masses of light quarks as the {\it effective} masses~\cite{Jegerlehner:1990uiq,Marciano:1993jd}, which have been adopted to derive the hadronic contributions to the vacuum polarization and are evaluated from the measurements of $R \equiv \sigma(e^+ + e^- \to {\rm hadrons})/\sigma(e^+ + e^- \to \mu^+ +\mu^-)$~\cite{Marciano:1983ss,Hollik:1988ii,Denner:1991kt}.
	
	In terms of the number densities of protons $N_p^{}$ and neutrons $N_n^{}$ in matter, the relative corrections to the NC potential are evaluated as $\Delta c_{\rm V,NC}^{} / c_{\rm V,NC}^{} \approx 0.062 + 0.020 N^{}_p / N^{}_n$. For ordinary matter with $N^{}_p \approx N^{}_n$, the correction is about $8.2\%$, which is the same for all three neutrino flavors and does not affect flavor conversion.\footnote{The flavor-dependent corrections, which were first calculated in Ref.~\cite{Botella:1986wy}, are omitted here, since they are three orders of magnitude smaller than the flavor-independent ones.} For the CC potential, the relative correction $\Delta c^e_{\rm V, CC}/c^e_{\rm V,CC}$ is about $5.8\%$, which is relevant for neutrino oscillations. For antineutrinos, the corrections to the vector-type couplings are the same as those for neutrinos due to the crossing symmetry of the amplitudes, and one also needs to change the overall sign of the matter potential as in the tree-level case.

    In practice, the tree-level matter potential is conventionally expressed by using the Fermi coupling constant $G_{\mu}^{} \approx 1.166 \times 10^{-5}~{\rm GeV}^{-2}$, extracted from measurements of muon lifetime, namely, ${\cal V}_{\rm CC}^{} = \sqrt{2} G_{\mu}^{} N_e^{} c_{\rm V,CC}^e$ for the CC potential, instead of using the on-shell parameters $\left\{\alpha,m_W^{},m_Z^{}\right\}$ in Eq.~(\ref{eq:Veff}). At one-loop level, the relation between these two sets of parameters is given by
    \begin{eqnarray}
        \sqrt{2} G_\mu^{} = \frac{\pi \alpha m_Z^2}{m_W^2 \left(m_Z^2 - m_W^2\right)} (1+\Delta r) \;,
    \end{eqnarray}
    where $\Delta r \approx 3.8\%$ is the one-loop correction to muon decay~\cite{Sirlin:1980nh,Denner:1991kt,Denner:2019vbn}. In this case, the matter potential at tree level is larger than that expressed in Eq.~(\ref{eq:Veff}) with $\left\{\alpha,m_W^{},m_Z^{}\right\}$ as input parameters, since $G_\mu^{}$ itself already includes the effects of radiative corrections through $\Delta r$. On the other hand, the matter potential at one-loop level remains unchanged, since all radiative corrections have already been considered. The corrected CC potential then reads
    \begin{equation}
        \widehat{\cal V}_{\rm CC}^{} = \sqrt{2} G_\mu^{\rm LO} N_e^{} \widehat{c}_{\rm V,CC}^{e} \simeq \sqrt{2} G_\mu^{\rm NLO} N_e^{} c_{\rm V,CC}^{e} \left(1 + \Delta c_{\rm V,CC}^e - \Delta r \right),
    \end{equation}
    where $G_\mu^{\rm LO} = G_\mu^{} \left(1 - \Delta r\right)$ is the Fermi coupling constant evaluated at tree level, while $G_\mu^{\rm NLO} = G_\mu$ takes into account one-loop effects. Therefore, the relative correction to the CC potential decreases to $\Delta c_{\rm V,CC}^e - \Delta r \approx 2.0\%$, when expressed using the Fermi coupling constant $G^{}_\mu$.

    \subsection{Matter Density Profile and Its Uncertainties}

    Another quantity that affects the matter potential is the electron number density in matter. For a constant density, the matter potential itself has also a constant value. However, in experiments, the matter density usually changes during propagation of neutrinos. Therefore, the probability in Eq.~(\ref{eq:numu_to_nune}) is only valid for a small domain of the matter parameter $a$, in which the matter density $\rho$ can be approximated as a constant. Thus, the complete neutrino oscillation probability is proportional to the absolute value squared of the total evolution operator, which is the ordered product of all individual evolution operators with one evolution operator for each given domain~\cite{Ohlsson:1999um}.
    
    In general, the Earth's matter density profile always has uncertainties of the order of a few percent. Such uncertainties will have an impact on the calculation of the neutrino oscillation probabilities, especially close to the resonance region, {\em i.e.}, where $\Delta_{31} \simeq a L$ holds in Eq.~(\ref{eq:numu_to_nune})~\cite{Jacobsson:2001zk,Jacobsson:2002nb,Ohlsson:2003ip}. Although it is hard to realize the resonance in reality with a neutrino energy of $E_\nu^{} \in [2,4]~{\rm GeV}$ for accelerator neutrino experiments, such uncertainties in the matter density will still affect the sensitivity to determine the sign of $\Delta m_{31}^2$ and measure the value of $\delta_{\rm CP}^{}$ (see, {\em e.g.}, Refs.~\cite{Shan:2001br,Geller:2001ix,Huber:2002mx,Shan:2002px,Kelly:2018kmb,Ghosh:2022bqj} for previous studies on this effect). For DUNE, a constant matter density profile is usually chosen with $\rho_{\rm avg}^{} = 2.848~{\rm g}/{\rm cm}^{3}$ and a $5\%$ uncertainty to represent the average value of matter density (with uncertainties) along the baseline. Another matter density profile is the Shen-Ritzwoller profile~\cite{Shen:2016kxw}, which is able to describe the matter density along the baseline of DUNE more precisely and has been used in the simulation setup given in Ref.~\cite{DUNE:2021cuw}. In this work, we adopt a more conservative uncertainty of $2\%$, but the actual uncertainty can even be smaller ({\em e.g.}, ``of the order of $1\%$'' \cite{Roe:2017zdw}). 

    With the above information on the matter potential, we are ready to calculate neutrino oscillation probabilities and simulate neutrino events for future long-baseline accelerator neutrino experiments. Meanwhile, the impact of one-loop effects on the matter potential and the uncertainties in the matter density profile will also be taken into account.

	\section{Numerical Analysis of Neutrino Oscillations in DUNE}
	\label{sec:num}

    In this section, we examine the effects of one-loop corrections in neutrino oscillation experiments. As an example, we present a case study for DUNE, which is a long-baseline neutrino oscillation experiment currently under construction~\cite{DUNE:2020ypp}. Our numerical simulations are performed with \texttt{GLoBES}~\cite{Huber:2004ka,Huber:2007ji}. We assume the experimental configuration used in the DUNE Technical Design Report~\cite{DUNE:2021cuw}. This configuration assumes DUNE to collect data for 6.5~years in neutrino mode and 6.5~years in antineutrino mode. The baseline length is $L=1284.9~{\rm km}$ with an average matter density of $\rho^{}_{\rm avg} = 2.848~{\rm g} / {\rm cm}^3$. For the matter density profile, we consider both the approximation of constant matter density and the Shen-Ritzwoller (matter density) profile following Ref.~\cite{Roe:2017zdw}. The values of the neutrino oscillation parameters are adopted from the global best-fit values reported in \texttt{NuFIT~6.0}~\cite{Esteban:2024eli, NuFIT:6.0}. Unless otherwise stated, we consider the dataset that assumes normal neutrino mass ordering, {\em i.e.}, $\Delta m_{31}^2 > 0$, and includes the atmospheric neutrino data from the Super-Kamiokande experiment.

    Although observables are independent of renormalization schemes, it is important to pay attention to various input parameters in different schemes and to apply the correct relations between any two sets of parameters. If the on-shell scheme with the input parameters $\{\alpha, m^{}_W, m^{}_Z\}$ is used and the CC matter potential ${\cal V}^{}_{\rm CC}$ is calculated at tree level and $\widehat{\cal V}^{}_{\rm CC}$ at one-loop level, the relative correction to the matter potential is about $5.8\%$. However, in terms of the Fermi coupling constant $G^{}_\mu$, directly extracted from measurements of muon lifetime, this correction is $2.0\%$, since a part of the radiative correction has already been included in the input parameter $G^{}_\mu$. Since the matter potential is parametrized in terms of $G^{}_\mu$ in \texttt{GLoBES}, we only consider the $2.0\%$ correction in our simulations. Now, the one-loop effects are taken into account in the simulations as corrections to the matter density profile. Note that one-loop corrections to the NC potential $\mathcal{V}_{\rm NC}^{}$ are omitted, since $\mathcal{V}_{\rm NC}^{}$ does not contribute to neutrino oscillation probabilities.
    
\begin{figure}[!t]
		\centering
		\includegraphics[width=0.7\linewidth]{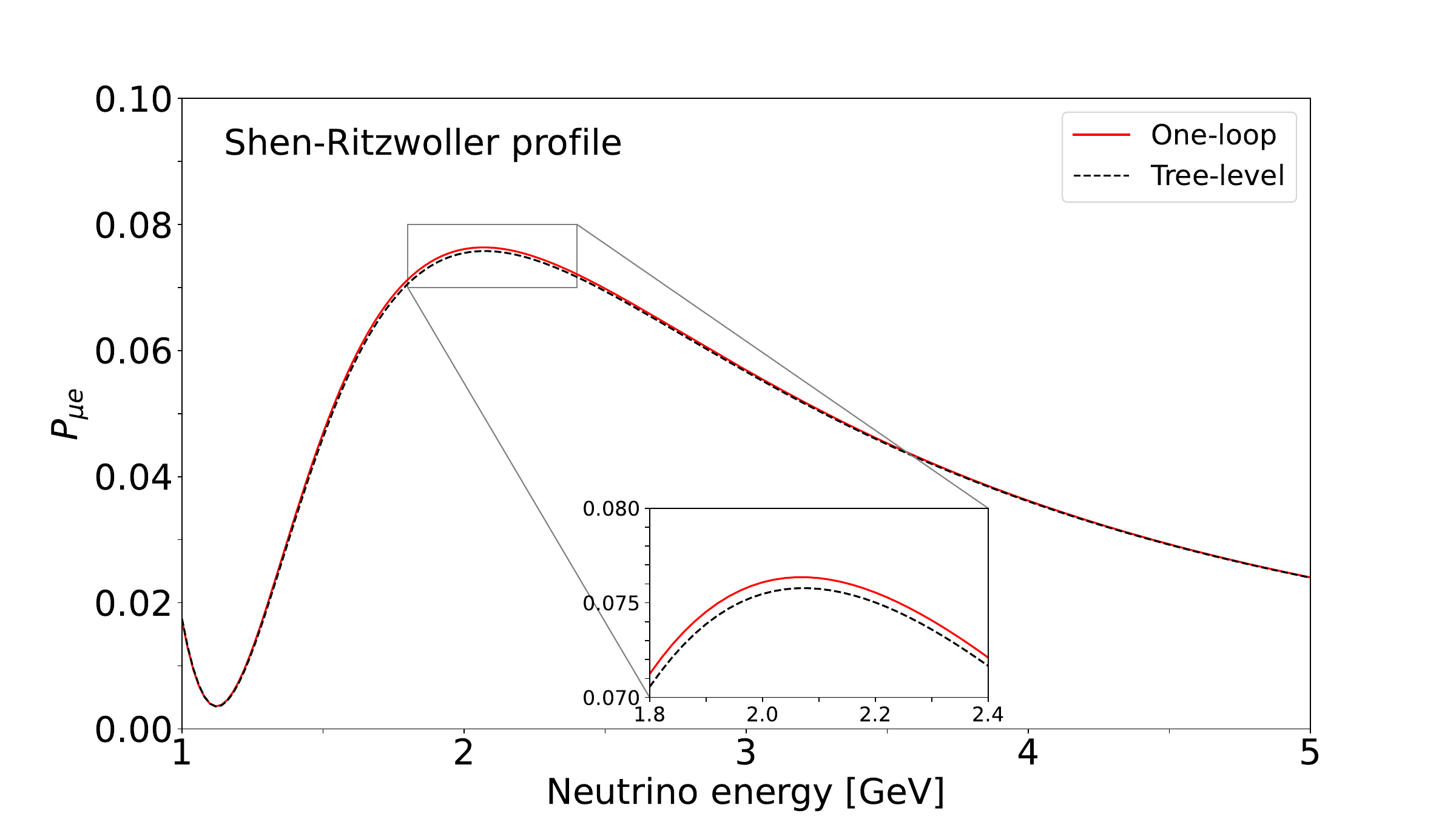} 
		\caption{The probability for $\nu_\mu^{} \rightarrow \nu_e^{}$ oscillations as the function of neutrino energy $E_\nu^{}$. The probability is shown for the setup of DUNE both at tree and one-loop levels for the Shen-Ritzwoller profile, while assuming the normal ordering for neutrino masses and a $2.0\%$ correction to the matter potential.}
		\label{fig:Probabilities}
	\end{figure}

    In Fig.~\ref{fig:Probabilities}, the neutrino oscillation probabilities for the $\nu_\mu^{} \rightarrow \nu_e^{}$ channel in DUNE are presented. The figure shows the neutrino oscillation probabilities for $E_\nu^{} \in [1, 5]~{\rm GeV}$ both at tree level (black dashed curve) and with one-loop corrections (red solid curve) by adopting the Shen-Ritzwoller profile. The one-loop effects are shown for the $2.0\%$ correction. It can be readily seen that the one-loop effects amount to an increase of about $\Delta P_{\mu e}^{} \equiv P_{\mu e}^{\rm NLO} - P_{\mu e}^{\rm LO} \simeq 0.0007$ near the first oscillation maximum, where $P_{\mu e}^{\rm NLO}$ and $P_{\mu e}^{\rm LO}$ are the probabilities computed at the tree and one-loop levels, respectively. In contrast, the difference between constant matter density and the Shen-Ritzwoller profile is found to be small. We have similarly computed the neutrino oscillation probability for the antineutrino channel $\overline{\nu}_\mu^{} \rightarrow \overline{\nu}_e^{} $. We find the one-loop effects on the antineutrino probability to be about $\Delta \overline{P}_{\mu e}^{} \equiv \overline{P}_{\mu e}^{\rm NLO} - \overline{P}_{\mu e}^{\rm LO} \simeq -0.0003$. In the case that the neutrino mass ordering were inverted, {\em i.e.}, $\Delta m_{31}^2 < 0$, then the corresponding one-loop effects would be $\Delta P_{\mu e}^{} \simeq -0.0003$ and $\Delta \overline{P}_{\mu e}^{} \simeq 0.0005$ instead.

\begin{figure}[!t]
		\centering
		\includegraphics[width=1.0\linewidth]{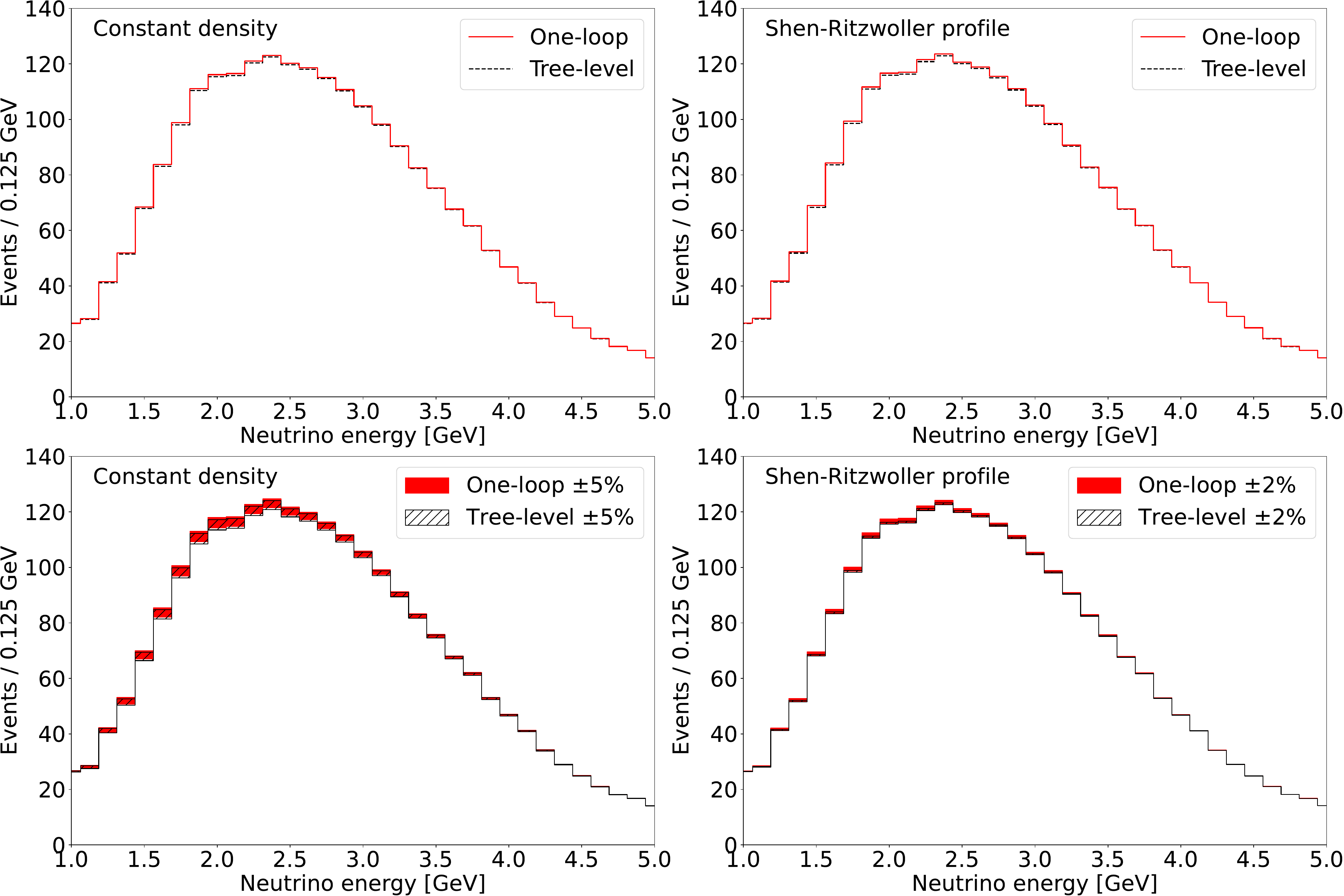} 
		\caption{The expected number of $\nu_e^{}$ events binned over the neutrino energies $[1,5]~{\rm GeV}$ in the setup of DUNE. Presented are the events at tree and one-loop levels for constant matter density (top-left) and the Shen-Ritzwoller profile (top-right), as well as the uncertainties associated with constant matter density (bottom-left) and the Shen-Ritzwoller profile (bottom-right). The events at one-loop level are shown for the $2.0\%$ correction.}
		\label{fig:Events}
	\end{figure}

    To see the difference between the tree-level and one-loop contributions, it is necessary to investigate effects of tree-level and one-loop contributions on the neutrino events that can be expected in DUNE. This effect is illustrated in Fig.~\ref{fig:Events}, where the expected number of $\nu_e^{}$ events is presented as a function of neutrino energies in the $E_\nu^{} \in [1,5]~{\rm GeV}$ range. The one-loop events are shown for the $2.0\%$ correction. The neutrino events include both the $\nu_\mu^{} \rightarrow \nu_e^{}$ signal and the corresponding backgrounds. The top-left and top-right panels show the binned $\nu_e^{}$ events that emerge from the $\nu_\mu^{} \rightarrow \nu_e^{}$ channel for constant matter density and the Shen-Ritzwoller profile, respectively. On the other hand, the bottom-left and bottom-right panels in Fig.~\ref{fig:Events} show the same neutrino events, but including also the uncertainties that are associated with constant matter density and the Shen-Ritzwoller profile.
    
     In Fig.~\ref{fig:Events}, the top-left and top-right panels show that the one-loop correction can produce noticeable effects in the $\nu_e^{}$ spectrum. When the one-loop effects are taken into account, the total number of $\nu_e^{}$ events is increased by 11 events for the $2.0\%$ correction in comparison to the tree-level contribution. Correspondingly, the Shen-Ritzwoller profile yields 8 neutrino events more than constant matter density. We have confirmed separately that a similar correction takes place in the $\overline{\nu}_e^{}$ spectrum, but with a negative sign. In case that the neutrino mass ordering were inverted, the effects in the $\nu_e^{}$ and $\overline{\nu}^{}_e$ spectra would be switched. 
    
    The difference between constant matter density and the Shen-Ritzwoller profile becomes evident in the bottom-left and bottom-right panels of Fig.~\ref{fig:Events}. When constant matter density is used in the calculation of the neutrino events (bottom-left panel), the tree-level and one-loop contributions are partially overlapping. On the other hand, the smaller uncertainty considered for the Shen-Ritzwoller profile (bottom-right panel) leads to a situation where the neutrino events computed with the one-loop corrections are clearly above those that are obtained at tree level. We have similarly investigated the effect of the one-loop corrections on the other neutrino oscillation channels that will be studied in DUNE. We have found that the one-loop corrections have a negligible effect for the $\nu_\mu^{}$ and $\overline{\nu}_\mu^{}$ events, whereas $\overline{\nu}_e^{}$ events are shifted downwards in a similar manner as what can be seen for the $\nu_e^{}$ events in Fig.~\ref{fig:Events}. It can therefore be expected that the one-loop corrections have an effect on the sensitivity to the neutrino mass ordering.
    
\begin{figure}[!t]
		\centering
		\includegraphics[width=1.0\linewidth]{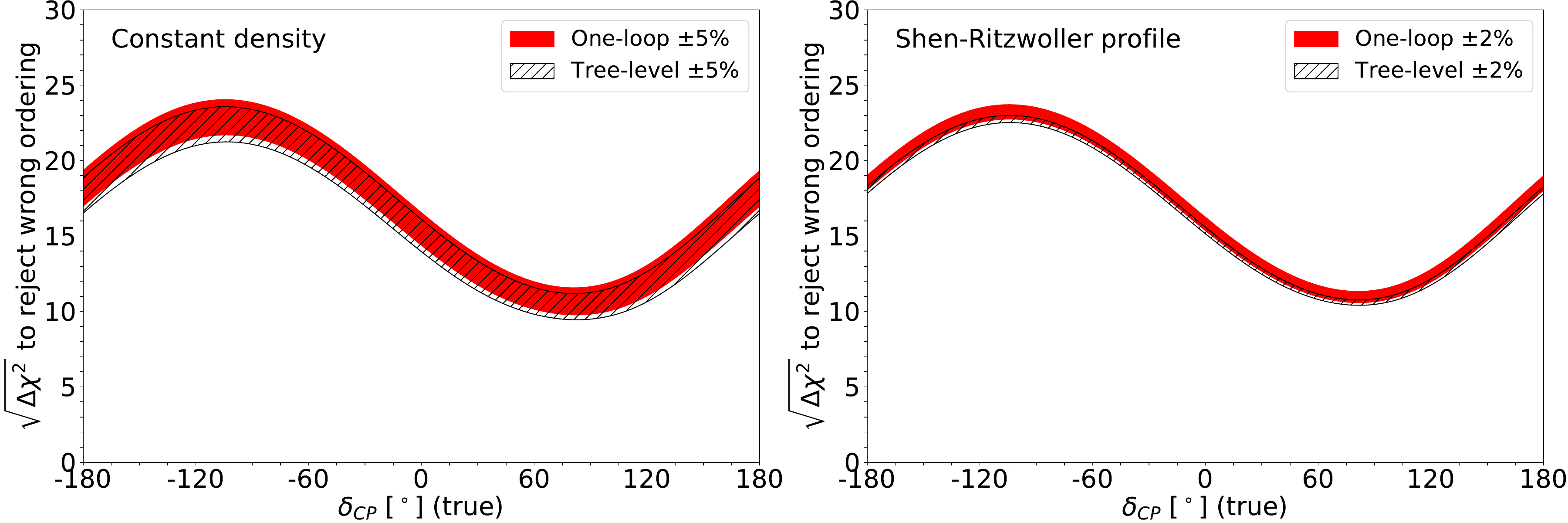} 
		\caption{Sensitivity to rule out the wrong mass ordering in DUNE. Left panel: Sensitivities obtained for constant matter density $\rho = 2.848~{\rm g}/{\rm cm}^3$. Right panel: Sensitivities obtained for the Shen-Ritzwoller profile. The widths of the bands correspond to the matter density uncertainties.}
		\label{fig:Sensitivity}
	\end{figure}

    Next, we examine the sensitivity to the neutrino mass ordering in DUNE. In Fig.~\ref{fig:Sensitivity}, the expected sensitivities to the neutrino mass ordering are presented for both constant matter density (left panel) and the Shen-Ritzwoller profile (right panel). The sensitivity to rule out the wrong neutrino mass ordering is shown as a function of the possible true values of $\delta_{\rm CP}^{}$. In both panels of Fig.~\ref{fig:Sensitivity}, we assume the true neutrino mass ordering to be NO. The sensitivity to the neutrino mass ordering is given by $\sqrt{\Delta \chi^2} = \sqrt{\chi^2_{\rm IO}}$, which has been computed by varying the test values of the neutrino oscillation parameters $\theta_{13}^{}, \theta_{23}^{}, \delta_{\rm CP}^{}$, and $\Delta m_{31}^2$ without priors. One may similarly obtain the sensitivity for the case where the true neutrino mass ordering is assumed to be IO. In that case, the sensitivity is calculated with $\sqrt{\Delta \chi^2} = \sqrt{\chi^2_{\rm NO}}$. For $\chi^2_{\rm IO}$, the variation of $\Delta m_{31}^2$ is carried out for the values of $\Delta m_{31}^2 < 0$, whereas for $\chi^2_{\rm NO}$, the variation is performed for the values of $\Delta m_{31}^2 > 0$. We keep the parameters $\theta_{12}^{}$ and $\Delta m_{21}^2$ fixed at their best-fit values~\cite{Esteban:2024eli,NuFIT:6.0}, as they have been determined with high precision from solar neutrino data and will be measured at the sub-percent level in JUNO~\cite{JUNO:2022mxj, Capozzi:2025wyn}. The effect of the matter density uncertainty is illustrated by varying the true value of $\rho$ within its uncertainties. This variation gives rise to the bandwidths that are visible in Fig.~\ref{fig:Sensitivity}. The systematic uncertainties are taken into account using the methods introduced in Ref.~\cite{DUNE:2021cuw}. More information on the systematic uncertainties can be found in Ref.~\cite{DUNE:2020ypp}.

    By studying the sensitivities to the neutrino mass ordering in Fig.~\ref{fig:Sensitivity}, it is observed that the one-loop effects correspond to an improvement of approximately $0.4\sigma$~CL for the $2.0\%$ correction. This improvement is obtained by marginalizing the sensitivities for the true values of $\rho$ and $\delta_{\rm CP}^{}$. It corresponds to the smallest change in the sensitivity over the true values of $\delta_{\rm CP}^{}$, while assuming the true neutrino mass ordering to be NO. Similarly, we obtain $0.4\sigma$~CL, when the true neutrino mass ordering is IO. It is observed for both neutrino mass orderings that the one-loop effect is the same for both constant matter density and the Shen-Ritzwoller profile. The one-loop effect is also visible when the uncertainties pertaining to $\rho$ are taken into account. When the uncertainty in the matter density is considered for constant matter density, we find the tree-level and one-loop sensitivities to become partially overlapped. In that case, the effect of the one-loop correction does not significantly differ from the tree-level sensitivities due to the matter density uncertainty. When the Shen-Ritzwoller profile is used instead, the difference between the tree-level and one-loop contributions to the neutrino mass ordering becomes observable. It is observed that a minor overlap occurs for the tree-level and one-loop sensitivities. It can therefore be inferred that the one-loop corrections to the matter potential lead to an effect that is difficult to discern from the uncertainties related to constant matter. However, as the matter density uncertainty becomes smaller with more accurate profiles such as the Shen-Ritzwoller profile, the effect of the one-loop corrections becomes more relevant.
    
     \begin{figure}[!t]
		\centering
		\includegraphics[width=0.7\linewidth]{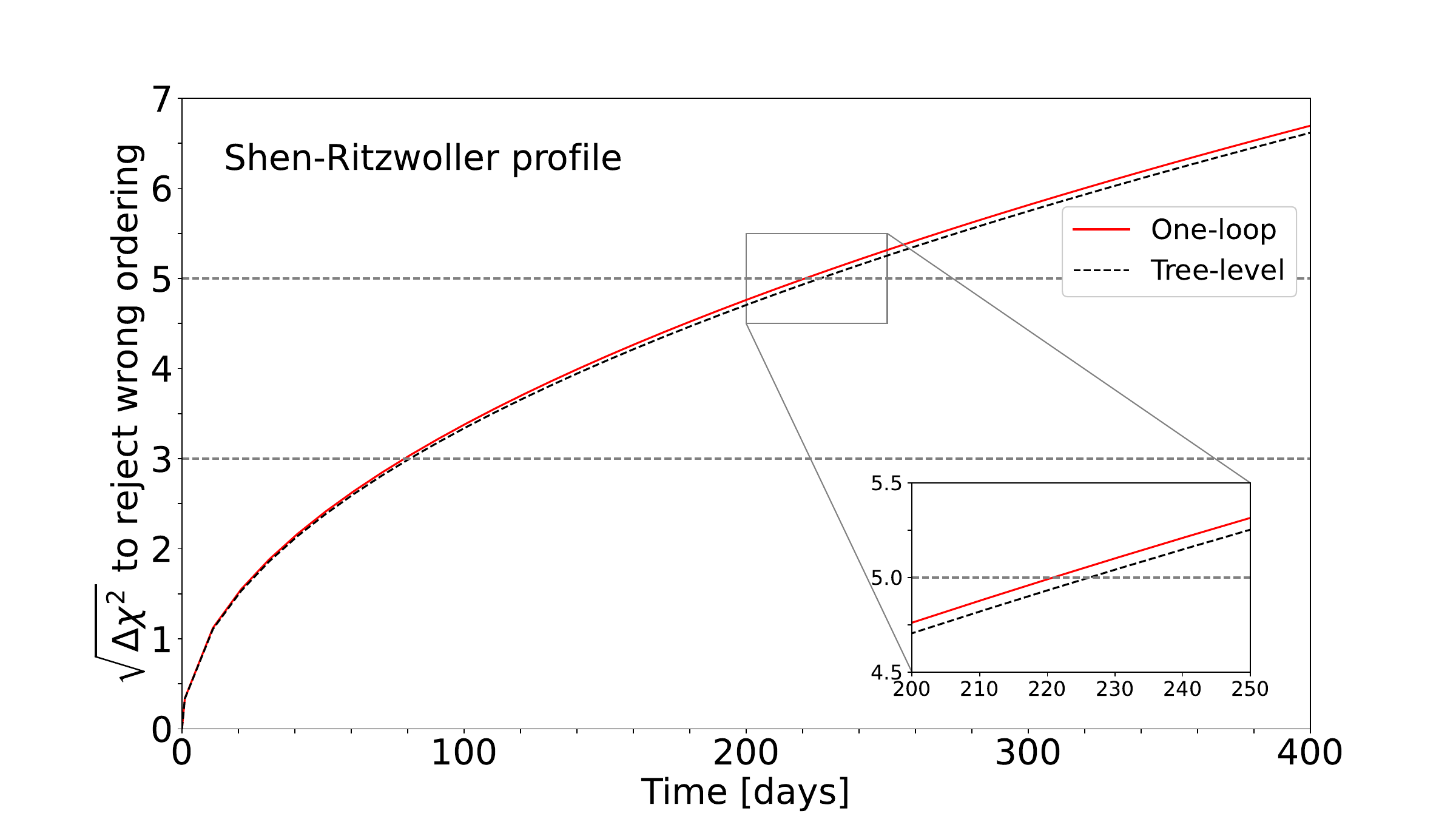} 
		\caption{Sensitivity to the neutrino mass ordering in DUNE as the function of days the experiment collects data. The sensitivity is presented for the tree-level and one-loop calculations, showing the $2.0\%$ correction. The effect of the matter density uncertainty is taken into account.}
		\label{fig:TimeDependence}
	\end{figure}

    The one-loop corrections to the matter potential may potentially lead to an earlier discovery of the neutrino mass ordering in DUNE. This prospect is illustrated in Fig.~\ref{fig:TimeDependence}, where the sensitivity to the mass ordering in DUNE is presented as the function of days during which the experiment collects data. As before, the $\sqrt{\Delta \chi^2} = \sqrt{\chi^2_{\rm IO}}$ distribution corresponds to the number of standard deviations at which the wrong neutrino mass ordering can be excluded in DUNE, when the true neutrino mass ordering is assumed to be NO. In this case, the minimization of $\chi^2_{\rm IO}$ also includes $\rho$. The corresponding $3\sigma$ and $5\sigma$~CL limits are indicated by the dashed gray lines. Figure~\ref{fig:TimeDependence} shows that the sensitivity to exclude IO in DUNE reaches $5\sigma $~CL approximately 5 days earlier due to the one-loop effects, when the correction is assumed to be $2.0\%$. The difference is calculated with respect to the standard tree-level matter potential. The true value of $\delta_{\rm CP}^{}$ is assumed to be $\delta_{\rm CP}^{} = 212^\circ,$ which corresponds to the global best-fit value according to \texttt{NuFit 6.0}~\cite{NuFIT:6.0}. We have similarly computed the sensitivity to the neutrino mass ordering assuming the true value of $\delta_{\rm CP}^{}$ to be $270^\circ$ ($180^\circ$). In that case, the neutrino mass ordering would be discovered at $5\sigma$~CL about 4 days (9 days) earlier. These results have been computed for the Shen-Ritzwoller profile. We have obtained similar results for constant density.

    In addition to the neutrino mass ordering, we investigated the effect of one-loop corrections on the sensitivities to the leptonic CP violation in DUNE. We specifically examined the effect on the CP violation discovery potential and the precision at which $\delta_{\rm CP}^{}$ can be determined in DUNE. For both cases, we confirmed that the one-loop corrections lead to insignificant changes in the physics reach of DUNE. This result is also expected, since the sensitivity to CP violation is mostly independent of matter effects.

	\section{Summary}
	\label{sec:sum}

	In this work, we have investigated the one-loop effects on the matter potential in the next-generation long-baseline neutrino experiment DUNE. It is found that the one-loop effects can lead to either a $5.8\%$ or $2.0\%$ correction to the charged-current potential depending on the choice of input parameters. For the first time, we show these effects on the physics reach of DUNE.

    Investigations at the neutrino oscillation probability level have revealed that one-loop effects lead to a noticeable increase in the electron neutrino appearance channel $\nu_\mu^{} \to \nu_e^{}$. In contrast, the probability for the corresponding antineutrino channel $\overline{\nu}_\mu^{} \to \overline{\nu}_e^{}$ decreases due to one-loop effects. We have confirmed this behavior also at the event level, where the expected $\nu_e^{}$ events in DUNE increase by 11 events assuming the normal ordering for neutrino masses. The corresponding effect for the inverted ordering is decreasing $\nu_e^{}$ events and increasing $\overline{\nu}_e^{}$ events. In both cases, the one-loop effects on the muon neutrino disappearance channel $\nu_\mu^{} \to \nu_\mu^{}$ and the antineutrino channel $\overline{\nu}_\mu^{} \to \overline{\nu}_\mu^{}$ are found to be too small to be observed. 
    
    In general, it is found that the one-loop effects lead to higher sensitivities to the neutrino mass ordering in DUNE. We have found that the sensitivity increases by about $0.4\sigma$~CL for both the normal ordering and the inverted ordering when the $2.0\%$ one-loop correction to the matter potential is considered. This result has been obtained after marginalizing the effects of the matter density uncertainty. This correction is present regardless of the true value of the Dirac CP-violating phase $\delta_{\rm CP}^{}$. We have furthermore determined that the $2.0\%$ one-loop correction leads to the neutrino mass ordering to be resolved at $5\sigma$~CL roughly 4--9 days earlier than for the tree-level prediction, depending on the value of $\delta_{\rm CP}^{}$. Finally, we have found no significant changes in the sensitivities to CP violation in the lepton sector due to one-loop effects. Although one-loop corrections to the matter potential are at the level of $2.0\%$ if the Fermi coupling constant $G^{}_\mu$ is chosen as input, they arise from the standard electroweak interactions and should be incorporated in future determination of the neutrino oscillation parameters and new-physics searches in long-baseline accelerator neutrino experiments.

	\section*{Acknowledgements}
	
	This work was supported in part by the National Natural Science Foundation of China under grant No.~12475113, by the CAS Project for Young Scientists in Basic Research (YSBR-099), and by the Scientific and Technological Innovation Program of IHEP under grant No.~E55457U2.

	\bibliographystyle{elsarticle-num}
	\bibliography{ref_MSW_pheno}

\begin{thebibliography}{10}
\expandafter\ifx\csname url\endcsname\relax
  \def\url#1{\texttt{#1}}\fi
\expandafter\ifx\csname urlprefix\endcsname\relax\def\urlprefix{URL }\fi
\expandafter\ifx\csname href\endcsname\relax
  \def\href#1#2{#2} \def\path#1{#1}\fi

\bibitem{ParticleDataGroup:2024cfk}
S.~Navas, et~al., {Review of Particle Physics}, Phys. Rev. D 110 (2024) 030001.

\bibitem{Xing:2020ijf}
Z.-z. Xing, {Flavor structures of charged fermions and massive neutrinos},
  Phys. Rept. 854 (2020) 1--147.
\newblock \href {http://arxiv.org/abs/1909.09610} {\path{arXiv:1909.09610}}.

\bibitem{JUNO:2015zny}
F.~An, et~al., {Neutrino physics with JUNO}, J. Phys. G 43 (2016) 030401.
\newblock \href {http://arxiv.org/abs/1507.05613} {\path{arXiv:1507.05613}}.

\bibitem{DUNE:2020ypp}
B.~Abi, et~al., {Deep Underground Neutrino Experiment (DUNE), Far Detector
  Technical Design Report, Volume II: DUNE Physics} (2 2020).
\newblock \href {http://arxiv.org/abs/2002.03005} {\path{arXiv:2002.03005}}.

\bibitem{DeSalas:2018rby}
P.~F. De~Salas, S.~Gariazzo, O.~Mena, C.~A. Ternes, M.~T\'ortola, {Neutrino
  Mass Ordering from Oscillations and Beyond: 2018 Status and Future
  Prospects}, Front. Astron. Space Sci. 5 (2018) 36.
\newblock \href {http://arxiv.org/abs/1806.11051} {\path{arXiv:1806.11051}}.

\bibitem{Wolfenstein:1977ue}
L.~Wolfenstein, {Neutrino oscillations in matter}, Phys. Rev. D 17 (1978)
  2369--2374.

\bibitem{Wolfenstein:1979ni}
L.~Wolfenstein, {Neutrino oscillations and stellar collapse}, Phys. Rev. D 20
  (1979) 2634--2635.

\bibitem{Mikheyev:1985zog}
S.~P. Mikheyev, A.~Y. Smirnov, {Resonance enhancement of oscillations in matter
  and solar neutrino spectroscopy}, Sov. J. Nucl. Phys. 42 (1985) 913--917.

\bibitem{Mikheev:1986wj}
S.~P. Mikheyev, A.~Y. Smirnov, {Resonant Amplification of $\nu$ Oscillations in
  Matter and Solar-Neutrino Spectroscopy}, Nuovo Cim. C 9 (1986) 17--26.

\bibitem{Huang:2023nqf}
J.~Huang, S.~Zhou, {Mikheyev-Smirnov-Wolfenstein matter potential at the
  one-loop level in the Standard Model}, Phys. Rev. D 108 (2023) 093010.
\newblock \href {http://arxiv.org/abs/2307.04685} {\path{arXiv:2307.04685}}.

\bibitem{JUNO:2022mxj}
A.~Abusleme, et~al., {Sub-percent precision measurement of neutrino oscillation
  parameters with JUNO}, Chin. Phys. C 46 (2022) 123001.
\newblock \href {http://arxiv.org/abs/2204.13249} {\path{arXiv:2204.13249}}.

\bibitem{Capozzi:2025wyn}
F.~Capozzi, W.~Giar\`e, E.~Lisi, A.~Marrone, A.~Melchiorri, A.~Palazzo,
  {Neutrino masses and mixing: Entering the era of subpercent precision} (3
  2025).
\newblock \href {http://arxiv.org/abs/2503.07752} {\path{arXiv:2503.07752}}.

\bibitem{Bahcall:1995mm}
J.~N. Bahcall, M.~Kamionkowski, A.~Sirlin, {Solar neutrinos: Radiative
  corrections in neutrino-electron scattering experiments}, Phys. Rev. D 51
  (1995) 6146--6158.
\newblock \href {http://arxiv.org/abs/astro-ph/9502003}
  {\path{arXiv:astro-ph/9502003}}.

\bibitem{DUNE:2021cuw}
B.~Abi, et~al., {Experiment Simulation Configurations Approximating DUNE TDR}
  (3 2021).
\newblock \href {http://arxiv.org/abs/2103.04797} {\path{arXiv:2103.04797}}.

\bibitem{Shen:2016kxw}
W.~Shen, M.~H. Ritzwoller, {Crustal and uppermost mantle structure beneath the
  United States}, J. Geophys. Res. Solid Earth 121 (2016) 4306--4342.

\bibitem{Pontecorvo:1957cp}
B.~Pontecorvo, {Mesonium and Antimesonium}, Sov. Phys. JETP 6 (1958) 429--431.

\bibitem{Maki:1962mu}
Z.~Maki, M.~Nakagawa, S.~Sakata, {Remarks on the Unified Model of Elementary
  Particles}, Prog. Theor. Phys. 28 (1962) 870--880.

\bibitem{Cervera:2000kp}
A.~Cervera, A.~Donini, M.~B. Gavela, J.~J. Gomez~C{\'a}denas, P.~Hern{\'a}ndez,
  O.~Mena, S.~Rigolin, {Golden measurements at a neutrino factory}, Nucl. Phys.
  B 579 (2000) 17--55, [Erratum: Nucl.~Phys.~B 593 (2001) 731--732].
\newblock \href {http://arxiv.org/abs/hep-ph/0002108}
  {\path{arXiv:hep-ph/0002108}}.

\bibitem{Freund:2001pn}
M.~Freund, {Analytic approximations for three neutrino oscillation parameters
  and probabilities in matter}, Phys. Rev. D 64 (2001) 053003.
\newblock \href {http://arxiv.org/abs/hep-ph/0103300}
  {\path{arXiv:hep-ph/0103300}}.

\bibitem{Akhmedov:2004ny}
E.~K. Akhmedov, R.~Johansson, M.~Lindner, T.~Ohlsson, T.~Schwetz, {Series
  expansions for three-flavor neutrino oscillation probabilities in matter},
  JHEP 04 (2004) 078.
\newblock \href {http://arxiv.org/abs/hep-ph/0402175}
  {\path{arXiv:hep-ph/0402175}}.

\bibitem{Nunokawa:2007qh}
H.~Nunokawa, S.~J. Parke, J.~W.~F. Valle, {CP violation and neutrino
  oscillations}, Prog. Part. Nucl. Phys. 60 (2008) 338--402.
\newblock \href {http://arxiv.org/abs/0710.0554} {\path{arXiv:0710.0554}}.

\bibitem{Giunti:2007ry}
C.~Giunti, C.~W. Kim, {Fundamentals of Neutrino Physics and Astrophysics},
  {Oxford University Press}, 2007.

\bibitem{Xing:2011zza}
Z.-z. Xing, S.~Zhou, {Neutrinos in Particle Physics, Astronomy and Cosmology},
  {Springer Berlin, Heidelberg}, 2011.

\bibitem{Sirlin:1991fd}
A.~Sirlin, {Theoretical Considerations Concerning the $Z^0$ Mass}, Phys. Rev.
  Lett. 67 (1991) 2127--2130.

\bibitem{Sirlin:1991rt}
A.~Sirlin, {Observations concerning mass renormalization in the electroweak
  theory}, Phys. Lett. B 267 (1991) 240--242.

\bibitem{Pilaftsis:1997dr}
A.~Pilaftsis, {Resonant $CP$ violation induced by particle mixing in transition
  amplitudes}, Nucl. Phys. B 504 (1997) 61--107.
\newblock \href {http://arxiv.org/abs/hep-ph/9702393}
  {\path{arXiv:hep-ph/9702393}}.

\bibitem{Jegerlehner:1990uiq}
F.~Jegerlehner, {Renormalizing the standard model}, Conf. Proc. C 900603 (1990)
  476--590.

\bibitem{Marciano:1993jd}
W.~J. Marciano, {Spin and precision electroweak physics}, in: {21st Annual SLAC
  Summer Institute on Particle Physics: Spin Structure in High-energy Processes
  (School: 26 Jul--3 Aug 1993, Topical Conference: 4--6 Aug 1993) (SSI 93)},
  1993, pp. 35--56.

\bibitem{Marciano:1983ss}
W.~J. Marciano, A.~Sirlin, {Some general properties of the ${\cal O}(\alpha)$
  corrections to parity violation in atoms}, Phys. Rev. D 29 (1984) 75--88,
  [Erratum: Phys.~Rev.~D 31 (1985) 213].

\bibitem{Hollik:1988ii}
W.~F.~L. Hollik, {Radiative Corrections in the Standard Model and Their Role
  for Precision Tests of the Electroweak Theory}, Fortsch. Phys. 38 (1990)
  165--260.

\bibitem{Denner:1991kt}
A.~Denner, {Techniques for the Calculation of Electroweak Radiative Corrections
  at the One-Loop Level and Results for W-physics at LEP 200}, Fortsch. Phys.
  41 (1993) 307--420.
\newblock \href {http://arxiv.org/abs/0709.1075} {\path{arXiv:0709.1075}}.

\bibitem{Botella:1986wy}
F.~J. Botella, C.~S. Lim, W.~J. Marciano, {Radiative corrections to neutrino
  indices of refraction}, Phys. Rev. D 35 (1987) 896--901.

\bibitem{Sirlin:1980nh}
A.~Sirlin, {Radiative corrections in the ${\rm SU}(2)_L \times {\rm U}(1)$
  theory: A simple renormalization framework}, Phys. Rev. D 22 (1980) 971--981.

\bibitem{Denner:2019vbn}
A.~Denner, S.~Dittmaier, {Electroweak radiative corrections for collider
  physics}, Phys. Rept. 864 (2020) 1--163.
\newblock \href {http://arxiv.org/abs/1912.06823} {\path{arXiv:1912.06823}}.

\bibitem{Ohlsson:1999um}
T.~Ohlsson, H.~Snellman, {Neutrino oscillations with three flavors in matter:
  Applications to neutrinos traversing the Earth}, Phys. Lett. B 474 (2000)
  153--162, [Erratum: Phys.~Lett.~B 480 (2000) 419].
\newblock \href {http://arxiv.org/abs/hep-ph/9912295}
  {\path{arXiv:hep-ph/9912295}}.

\bibitem{Jacobsson:2001zk}
B.~Jacobsson, T.~Ohlsson, H.~Snellman, W.~Winter, {Effects of random matter
  density fluctuations on the neutrino oscillation transition probabilities in
  the Earth}, Phys. Lett. B 532 (2002) 259--266.
\newblock \href {http://arxiv.org/abs/hep-ph/0112138}
  {\path{arXiv:hep-ph/0112138}}.

\bibitem{Jacobsson:2002nb}
B.~Jacobsson, T.~Ohlsson, H.~Snellman, W.~Winter, {The effects of matter
  density uncertainties on neutrino oscillations in the Earth}, J. Phys. G 29
  (2003) 1873--1876.
\newblock \href {http://arxiv.org/abs/hep-ph/0209147}
  {\path{arXiv:hep-ph/0209147}}.

\bibitem{Ohlsson:2003ip}
T.~Ohlsson, W.~Winter, {Role of matter density uncertainties in the analysis of
  future neutrino factory experiments}, Phys. Rev. D 68 (2003) 073007.
\newblock \href {http://arxiv.org/abs/hep-ph/0307178}
  {\path{arXiv:hep-ph/0307178}}.

\bibitem{Shan:2001br}
L.-Y. Shan, B.-L. Young, X.-m. Zhang, {$CP$ violating neutrino oscillation and
  uncertainties in Earth matter density}, Phys. Rev. D 66 (2002) 053012.
\newblock \href {http://arxiv.org/abs/hep-ph/0110414}
  {\path{arXiv:hep-ph/0110414}}.

\bibitem{Geller:2001ix}
R.~J. Geller, T.~Hara, {Geophysical aspects of very long baseline neutrino
  experiments}, Nucl. Instrum. Meth. A 503 (2003) 187--191.
\newblock \href {http://arxiv.org/abs/hep-ph/0111342}
  {\path{arXiv:hep-ph/0111342}}.

\bibitem{Huber:2002mx}
P.~Huber, M.~Lindner, W.~Winter, {Superbeams vs.~neutrino factories}, Nucl.
  Phys. B 645 (2002) 3--48.
\newblock \href {http://arxiv.org/abs/hep-ph/0204352}
  {\path{arXiv:hep-ph/0204352}}.

\bibitem{Shan:2002px}
L.-Y. Shan, X.-M. Zhang, {Solar neutrino zenith angle distribution and
  uncertainty in Earth's matter density}, Phys. Rev. D 65 (2002) 113011.

\bibitem{Kelly:2018kmb}
K.~J. Kelly, S.~J. Parke, {Matter density profile shape effects at DUNE}, Phys.
  Rev. D 98 (2018) 015025.
\newblock \href {http://arxiv.org/abs/1802.06784} {\path{arXiv:1802.06784}}.

\bibitem{Ghosh:2022bqj}
M.~Ghosh, O.~Yasuda, {Effect of matter density in T2HK and DUNE}, Nucl. Phys. B
  989 (2023) 116142.
\newblock \href {http://arxiv.org/abs/2210.09103} {\path{arXiv:2210.09103}}.

\bibitem{Roe:2017zdw}
B.~Roe, {Matter density versus distance for the neutrino beam from Fermilab to
  Lead, South Dakota, and comparison of oscillations with variable and constant
  density}, Phys. Rev. D 95 (2017) 113004.
\newblock \href {http://arxiv.org/abs/1707.02322} {\path{arXiv:1707.02322}}.

\bibitem{Huber:2004ka}
P.~Huber, M.~Lindner, W.~Winter, {Simulation of long-baseline neutrino
  oscillation experiments with GLoBES (General Long Baseline Experiment
  Simulator)}, Comput. Phys. Commun. 167 (2005) 195--202.
\newblock \href {http://arxiv.org/abs/hep-ph/0407333}
  {\path{arXiv:hep-ph/0407333}}.

\bibitem{Huber:2007ji}
P.~Huber, J.~Kopp, M.~Lindner, M.~Rolinec, W.~Winter, {New features in the
  simulation of neutrino oscillation experiments with GLoBES 3.0 (General Long
  Baseline Experiment Simulator)}, Comput. Phys. Commun. 177 (2007) 432--438.
\newblock \href {http://arxiv.org/abs/hep-ph/0701187}
  {\path{arXiv:hep-ph/0701187}}.

\bibitem{Esteban:2024eli}
I.~Esteban, M.~C. Gonzalez-Garcia, M.~Maltoni, I.~Martinez-Soler, J.~P.
  Pinheiro, T.~Schwetz, {NuFit-6.0: updated global analysis of three-flavor
  neutrino oscillations}, JHEP 12 (2024) 216.
\newblock \href {http://arxiv.org/abs/2410.05380} {\path{arXiv:2410.05380}}.

\bibitem{NuFIT:6.0}
I.~Esteban, et~al., {NuFIT} 6.0, \url{http://www.nu-fit.org/} (2024).

\end{thebibliography}

\end{document}